\documentclass[preprint,english,showpacs,11pt,floatfix]{revtex4}

\usepackage{babel,amsmath,amssymb,dcolumn}
\usepackage[dvips]{graphics}
\usepackage{hyperref}

\usepackage{amsfonts}
\usepackage{amssymb}
\usepackage[dvips]{graphicx}
\usepackage{latexsym}

\usepackage{dcolumn}
\usepackage{bm}

\begin{document}

\title{Holographic explanation of wide-angle power
  correlation suppression in the Cosmic Microwave Background Radiation.}

\author{Zhuo-Yi Huang, Bin Wang}
\email{wangb@fudan.edu.cn} \affiliation{Department of Physics,
Fudan University, Shanghai 200433, People's Republic of China }

\author{Elcio Abdalla}
\email{eabdalla@fma.if.usp.br} \affiliation{Instituto de Fisica,
Universidade de Sao Paulo, C.P.66.318, CEP 05315-970, Sao Paulo,
Brazil}

\author{Ru-Keng Su}
\email{rksu@fudan.ac.cn} \affiliation{China Center of Advanced
Science and Technology (World Laboratory), P.B.Box 8730, Beijing
100080, People's Republic of China
\\Department of Physics, Fudan University, Shanghai 200433,
People's Republic of China }

\begin{abstract}
We investigate the question of the suppression of the CMB power spectrum
for the lowest multipoles in closed Universes. The intrinsic reason for a
lowest cutoff in closed Universes, connected with the discrete spectrum of
the wavelength, is shown not to be enough to explain observations.
We thus extend the holographic cosmic duality to closed
universes by relating the dark energy equation of state and the
power spectrum,  showing a suppression behavior which describes the
low $\ell$ features extremely well. We also explore the
possibility to disclose the nature of the dark energy from
the observed small $\ell$ CMB spectrum by employing the holographic idea.
\end{abstract}


\maketitle

In view of the example provided by the black hole entropy, an influential
holographic principle, relating the maximum number of degrees of
freedom in a volume to its boundary surface area, has been put
forward, attracting a lot of attention in the last decade
\cite{01}. The extension of the holographic principle to a general
cosmological setting was first addressed by Fischler and Susskind
\cite{02} and subsequently got modified by many authors
\cite{03,04,05,06,07}. The idea of holography is viewed as a
real conceptual change in our thinking about gravity \cite{08}. It
is interesting to note that holography implies a possible value
of the cosmological constant in a large class of universes
\cite{09}. In an inhomogeneous cosmology holography was also
realized as a useful tool to select physically acceptable models
\cite{06}. The idea of holography has further been applied to the
study of inflation and gives possible upper limits to the number of
e-folds \cite{10}. Recently, holography has again been proved
as an effective way to investigate dark energy \cite{11,12}.

The first year data from the Wilkinson Microwave Anisotropy Probe
(WMAP) manifest a spectacular agreement with the $\Lambda$CDM
model. However, as originally discovered by COBE, the measured
value of WMAP shows a suppression of the Cosmic Microwave
Background (CMB) anisotropy power on the largest angular scales
as compared to the standard model prediction. The explanation
of such missing wide-angle correlations was in the focus of much
discussion recently \cite{13,14,15,16,17}. Among them, an attempt to
relate the suppression of lower multipoles to the holographic
ideas is intriguing \cite{17}. It is also one more example of
applying holography in understanding cosmology.  The relation between the
equation of state of the dark energy and the features at low
multipoles in CMB power spectrum was built through a cosmic IR/UV
duality between the UV cutoff and a global infrared cutoff.

In this paper we would like to extend the discussion in \cite{17}
to a universe with spatial curvature. The tendency of preferring
a closed universe appeared in a suite of CMB experiments before
\cite{18}. The improved precision from WMAP provides further
confidence showing that a closed universe with positively curved
space is marginally preferred \cite{14,15,16,19}. In addition to
CMB, recently the spatial geometry of the universe was probed by
supernova measurements of the cubic correction  to the luminosity
distance \cite{20}, where a  closed universe is also marginally
favored. Closed universe models were used to explain the observed
lack of temperature correlations on large scale. In \cite{16}, a
simple geometrical model of a finite, positively curved space -
the Poincar\'{e} dodecahedral space, was used to account for the
WMAP's observations. In \cite{15}, it was shown that the low CMB
quadrupole amplitude can be reproduced if the primordial spectrum
truncates on scales comparable to the curvature scale. Our purpose
here is to relate the features in the power spectrum at low $\ell$
to the holographic idea in the framework of closed universe. We
will show that although a closed universe with finite size of
space itself imposes a cut-off on the comoving momentum and hence
suppressing the CMB spectrum at small $\ell$, the holographic
effect leads to the more reasonable position of the cutoff
$\ell_c$ in the multipole space comparing to the experiment. The
revealed correlation from holography between the dark energy and
the power spectrum will give us further constraints on the
equation of state of dark energy from the features at low
multipoles in the CMB power spectrum. Moreover, we shall compare the
results for the flat and the closed cases.


We will use coordinates for the metric of our universe
\begin{equation}
ds^2  = a^2 \left( \tau  \right)\left( { - d\tau ^2  + \frac{1}
{{\gamma ^2 }}\,\,d\textbf{x} \cdot d\textbf{x}} \right)
\label{metricline} 
\end{equation}
with $\gamma  = 1 + \frac{{Kr^2 }} {4}$, $ K =  - 1,\; + 1 $ and $
0 $ corresponding to the open, closed and flat universe,
respectively. For the flat universe, the CMB power spectrum in the
Sachs-Wolfe effect is
\begin{equation}
C_l  = \frac{{2}} {{25\pi}}\int_0^\infty  {\frac{{dk}}
{k}\mathcal{P_R} \left( k \right)j_l^2 \left( {k(\eta_0-\eta_{LS})} \right)}
\label{flatCl} 
\end{equation}
where $\mathcal{P_R} \propto k^{n - 1}$ is the curvature power
spectrum, $j_l \left( {k(\eta_0-\eta_{LS})} \right)$ the radial
dependence of the scalar harmonic function in the flat space and
$\eta_0 -\eta_{LS}$ the comoving distance to the last
scattering.  For the nonzero curvature space, the spherical
harmonic functions have been studied in \cite{21,22}. In a closed
universe the solution of the radial harmonic equation reads
\cite{22}
\begin{equation}
\Phi _\beta ^l  = \left( {\frac{{\pi M_\beta ^l }} {{2\beta ^2
\sin \xi }}} \right)^{{1 \mathord{\left/
 {\vphantom {1 2}} \right.
 \kern-\nulldelimiterspace} 2}} P_{ - {1 \mathord{\left/
 {\vphantom {1 2}} \right.
 \kern-\nulldelimiterspace} 2} + \beta }^{ - {1 \mathord{\left/
 {\vphantom {1 2}} \right.
 \kern-\nulldelimiterspace} 2} - l} \left( {\cos \xi } \right)
\label{closedradial} 
\end{equation}
where we defined the dimensionless parameter $\xi=K^{1/2}(\eta_0-\eta_{LS})$
and $ \beta ^2  = \frac{{k^2 }} {K} + 1 $ \cite{15} with $ \beta  = 3,4,
5... $ and $ \beta  > l $, such that we fulfill the requirement that $\Phi
_\beta ^l $ is single valued and satisfies a necessary periodic boundary
condition \cite{22}. The normalization factor is $ M_\beta ^l  =
\prod\limits_{n = 0}^l {\left( {\beta ^2  - n^2 } \right)}$. In a
closed universe the CMB power spectrum can be expressed as
\begin{equation}
C_l  = A\sum\limits_{\beta  \geqslant 3} {\frac{\beta } {{\beta ^2
- 1}}\mathcal{P_R} \left( \beta  \right)\left[ {\Phi _\beta ^l
\left( \eta_0-\eta_{LS} \right)} \right]^2 } \quad ,
\label{closedCl}  
\end{equation}
where the curvature power spectrum in the closed universe reads $
\mathcal{P_R} = \frac{{\left( {\beta ^2  - 4} \right)^2 }} {{\beta
^2 \left( {\beta ^2  - 1} \right)}}$ \cite{23}. Since the value of
$\beta$ is discrete due to the periodic boundary condition of the
harmonic function we have made the change $ \int {\frac{{dk}} {k}
\to \sum\limits_\beta  {\frac{\beta } {{\beta ^2  - 1}}} }$. From
Eq. (\ref{closedCl}) we see that since $\beta$ must be an integer,
starting from $3$ \cite{22}, implying  a lower bound for the
comoving momentum. 
If one takes the spatial curvature to zero, one has the solution of the
radial harmonic equation $\Phi_{\beta}^l=j_l(kr)$, thus the CMB spectrum
(\ref{closedCl}) will reduce to the flat case (\ref{flatCl}).
The discrete spectrum and the lower bound for the comoving
momentum are special properties brought by the nonzero spatial
curvature. It might be natural for a closed
universe to experience suppression in CMB spectrum at small
$\ell$s. However, we can see from Fig. 1 (the dashed line) that
such an intrinsic cutoff cannot be used to explain observations.
The position of the cutoff $\ell_c$ in the multipole space is too
small compared to the observational data \cite{24}.

Let us now  extend the cosmic duality model relating the
dark energy equation of state and the power spectrum put forward
in \cite{17} for the case of a closed universe.

Starting from the holographic idea relating the UV and IR cutoffs
suggested in \cite{cohen}, the dark energy duality in the universe
is $\rho_{\Lambda}=3d^2M_p^2 L^{-2}$, where $L$ is the largest IR
cutoff and $d$ is a free parameter; in the flat case $d\ge 1$ on general
grounds \cite{11,12}. Using the definitions $\Omega _\Lambda   = \frac{{\rho
_\Lambda  }} {{\rho _c }} $ and $ \rho _c  = 3M_p^2 H^2 $, we have, today,
$ L = \frac{d} {{\sqrt {\Omega _\Lambda ^0 } H_0 }} $.
Translating the IR cutoff $L$ into a cutoff at physical
wavelengths defined by $\lambda_c = 2L$ \cite{17}, we can compute the
smallest wave number today, $ k_c = \frac{\pi } {d}\sqrt
{\Omega _\Lambda ^0 } H_0 $. Expressing  such a cutoff in terms
of the index $\beta$, we have
\begin{equation}
\beta _c  = \sqrt {\frac{{k_c^2 }} {K} + 1}  = \sqrt {\frac{{\pi
^2 }} {{d^2 }}\frac{{\Omega _\Lambda ^0 }} {{\left( {\Omega
_{tot}^0  - 1} \right)}} + 1}  \label{betacutoff}\quad , 
\end{equation}
where $ K = H_0^2 \left( {\Omega _{tot}^0  - 1} \right)$ has been
used. We shall use $\beta_c$ to replace the cutoff $\beta=3$ brought by the
spatial curvature. In calculating the CMB spectrum (\ref{closedCl}), we 
will perform the summation for integer $\beta$s starting from
the smallest integer $\beta>\beta_c$. The discretization procedure applied in our work 
is directly from the periodic boundary condition in the closed universe, which is different from the 
discrete spectrum comes from the Neumann or Dirichlet boundary condition in \cite{17}.

The comoving distance to the last scattering follows from the
definition of comoving time
\begin{equation}
\eta _0  - \eta _{LS}  = \int_0^{z_{LS} } {dz'\frac{1} {{H\left(
{z'} \right)}}} \label{comovingtime}\quad . 
\end{equation}
In a closed universe whose main components are dark energy,
non-relativistic matter and curvature, we can write the Hubble parameter as
\begin{equation}
H^2 \left( z \right) = H_0^2 \left[ {\left( {\Omega _{tot}^0  -
\Omega _\Lambda ^0 } \right)\left( {1 + z} \right)^3  + \Omega
_\Lambda ^0 \left( {1 +
z} \right)^{3\left( {1 + w_0 } \right)} - \left( {\Omega _{tot}^0  - 1}
\right)\left( {1 + z} \right)^2 } \right] \label{components}\quad .
\end{equation} 
The distance to the last scattering depends on $w$. Thus the
relative position of the cutoff in the CMB spectrum depends on the
equation of state of dark energy. This exhibits the CMB/Dark
Energy cosmic duality, which was first realized in \cite{17}.

Given the experimental limits, $ \Omega _{tot}^0 = 1.02
$, $ \Omega _{\Lambda}^0 = 0.73 $, $w_{\Lambda}= - 1 $,
the position of the cutoff $\ell_c$ in the multipole space falls
at $ \ell_c  \sim 7$ (see the solid line in Fig. 1), which is
consistent with the observation \cite{24,13,14,15,16}.

\begin{figure}[!hbtp] \label{fig1}
\begin{center}
\includegraphics[width=10cm]{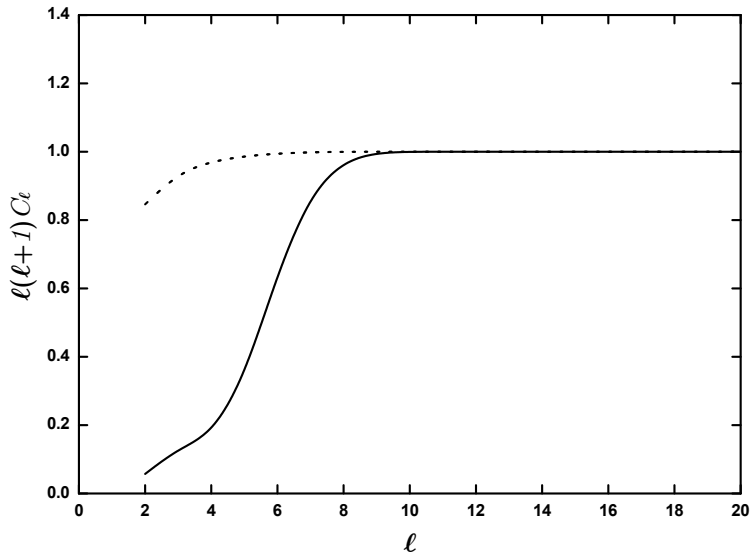}
\end{center}
\caption{The dashed line shows the suppression in the CMB spectrum
at small $l$ brought by the intrinsic reason due to the spatial
curvature. The solid line is the contribution by the holographic IR
cutoff. }
\end{figure}
\begin{figure}[!hbtp] \label{fig2}
\begin{center}
\includegraphics[width=10cm]{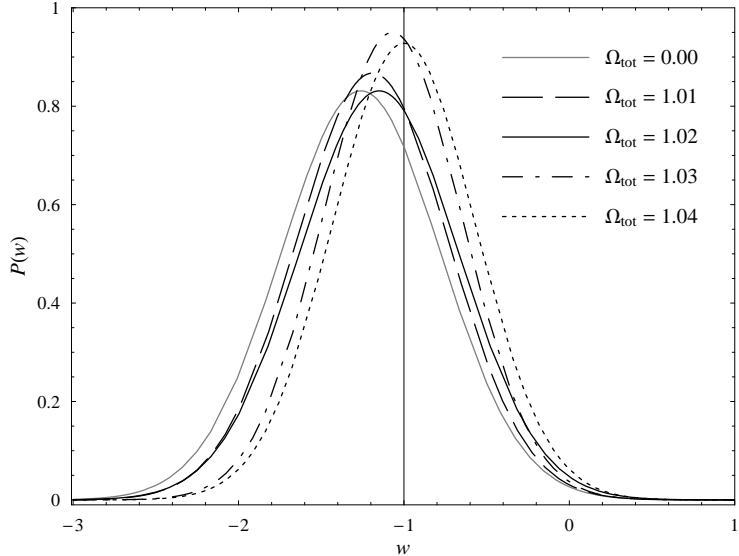}
\end{center}
\caption{ The distribution of the equation of state of dark energy
got by using cosmic duality from the suppression of CMB power
spectrum at $3<l_c<7$. The grey solid line is for the flat case, the
light grey dashed line for $\Omega_{tot}=1.01$, the black solid line
for $\Omega_{tot}=1.02$, the dash-dotted line in light grey for
$\Omega_{tot}=1.03$ and the dotted line in light grey for
$\Omega_{tot}=1.04$.}
\end{figure}

From the WMAP data, the statistically significant
suppression of the low multipole appears at the two first
multipoles corresponding to $l=2,3$. Combining data from WMAP and
other CMB experiments, the position of the cutoff $l_c$ in the
multipole space falls in the interval $3<l_c<7$. If we count on the 
the intrinsic reason provided by the natural cutoff implicit in
(\ref{closedCl}), brought by the spatial curvature, to explain
the observational features in the low $\ell$ CMB spectrum, we found that
for $\ell_c=2, \Omega_{tot}^0=1.02; \ell_c=3, \Omega_{tot}^0=1.04$ while
$\ell_c=4, \Omega_{tot}^0=1.08$ and $\Omega_{tot}^0$ continues to increase
for bigger values of $\ell_c$. Considering the observational constraint
$\Omega_{tot}^0=1.02^{+0.02}_{-0.02}$, it shows that the intrinsic reason
connected with the discrete spectrum of the wavelength for a lowest cutoff
in closed universes is not enough to explain the observation. The
generalized holographic idea can describe the low $\ell$ features well. 

\begin{figure}[!hbtp]
\begin{center}
\includegraphics[width=0.47\textwidth]{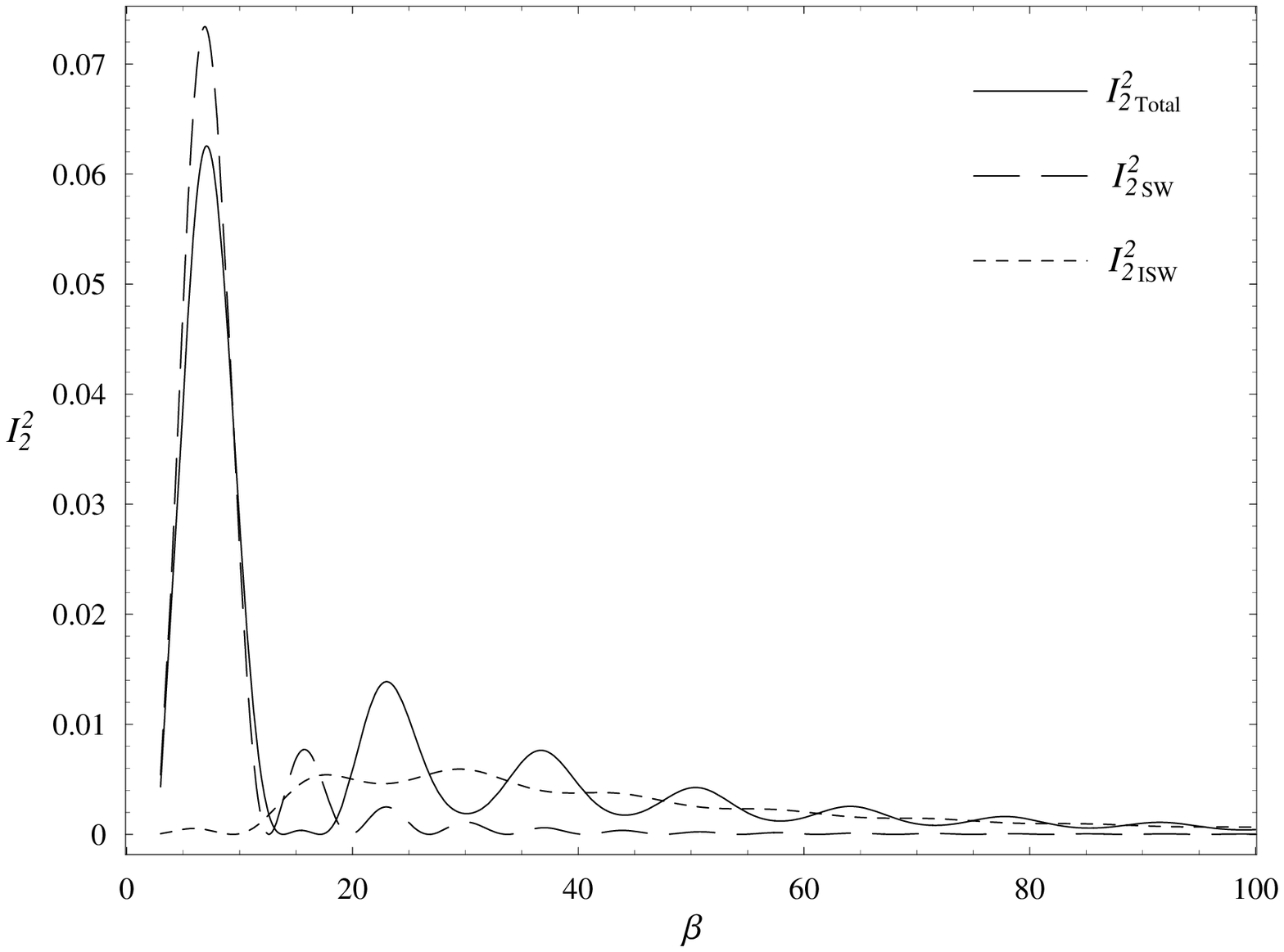}%
\hspace{0.05\textwidth}%
\includegraphics[width=0.47\textwidth]{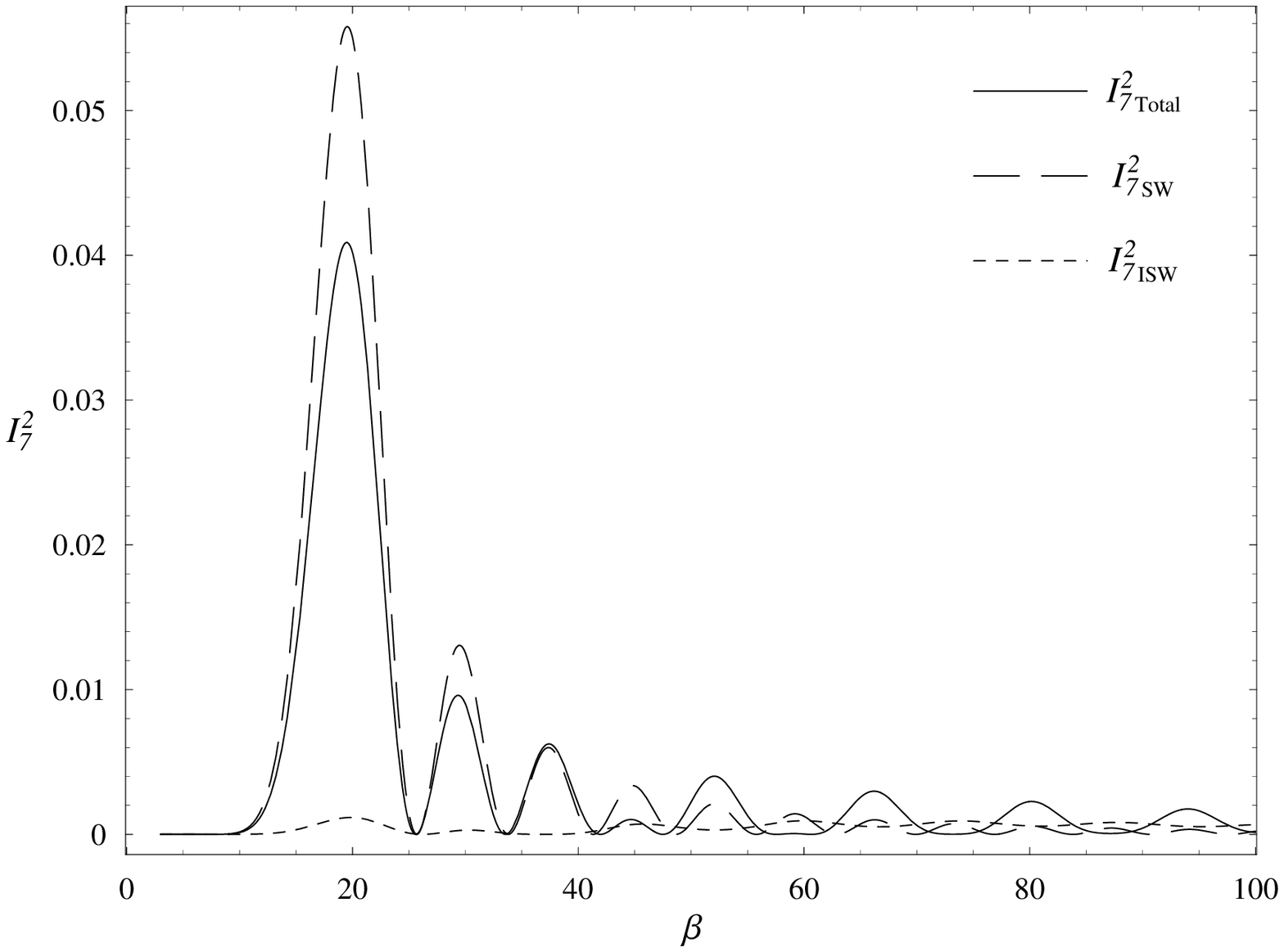}%
\end{center}
\caption{The Integrands of the SW and ISW contributions for $l=2,7$
respectively.}
\end{figure}
\begin{figure}[!hbtp]
\begin{center}
\includegraphics[width=0.4\textwidth]{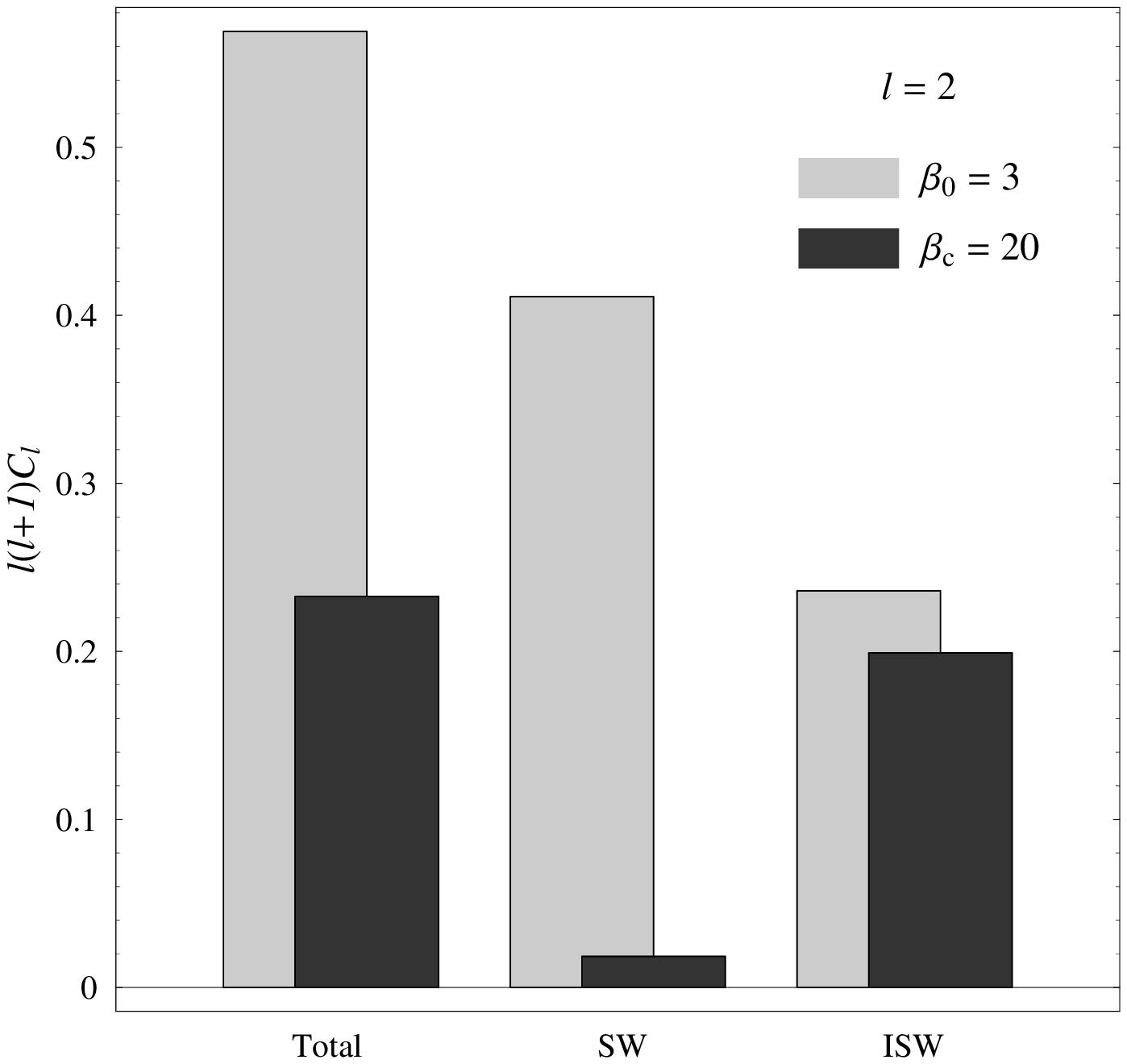}%
\hspace{0.05\textwidth}%
\includegraphics[width=0.4\textwidth]{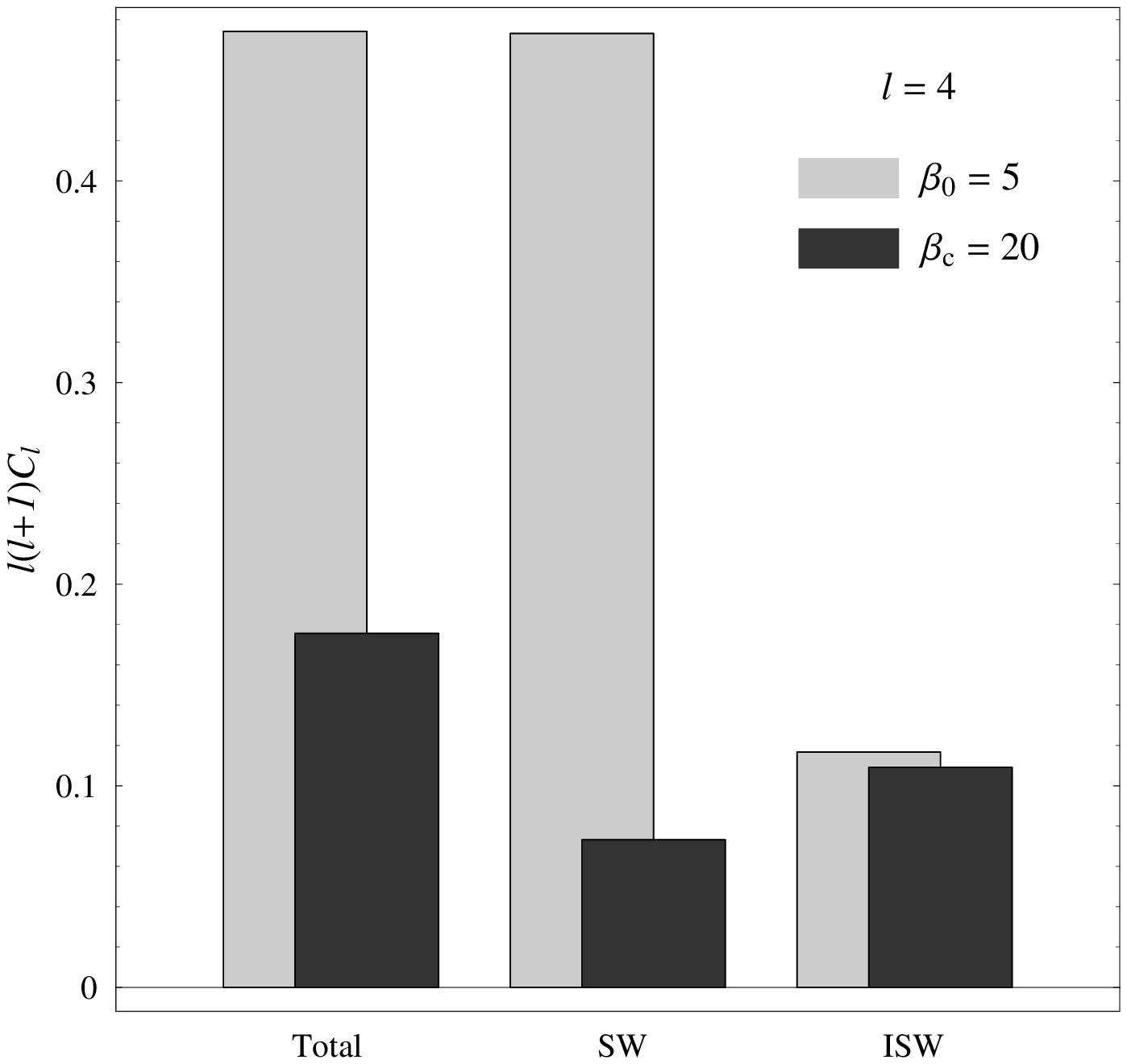}%
\hspace{0.05\textwidth}%
\includegraphics[width=0.4\textwidth]{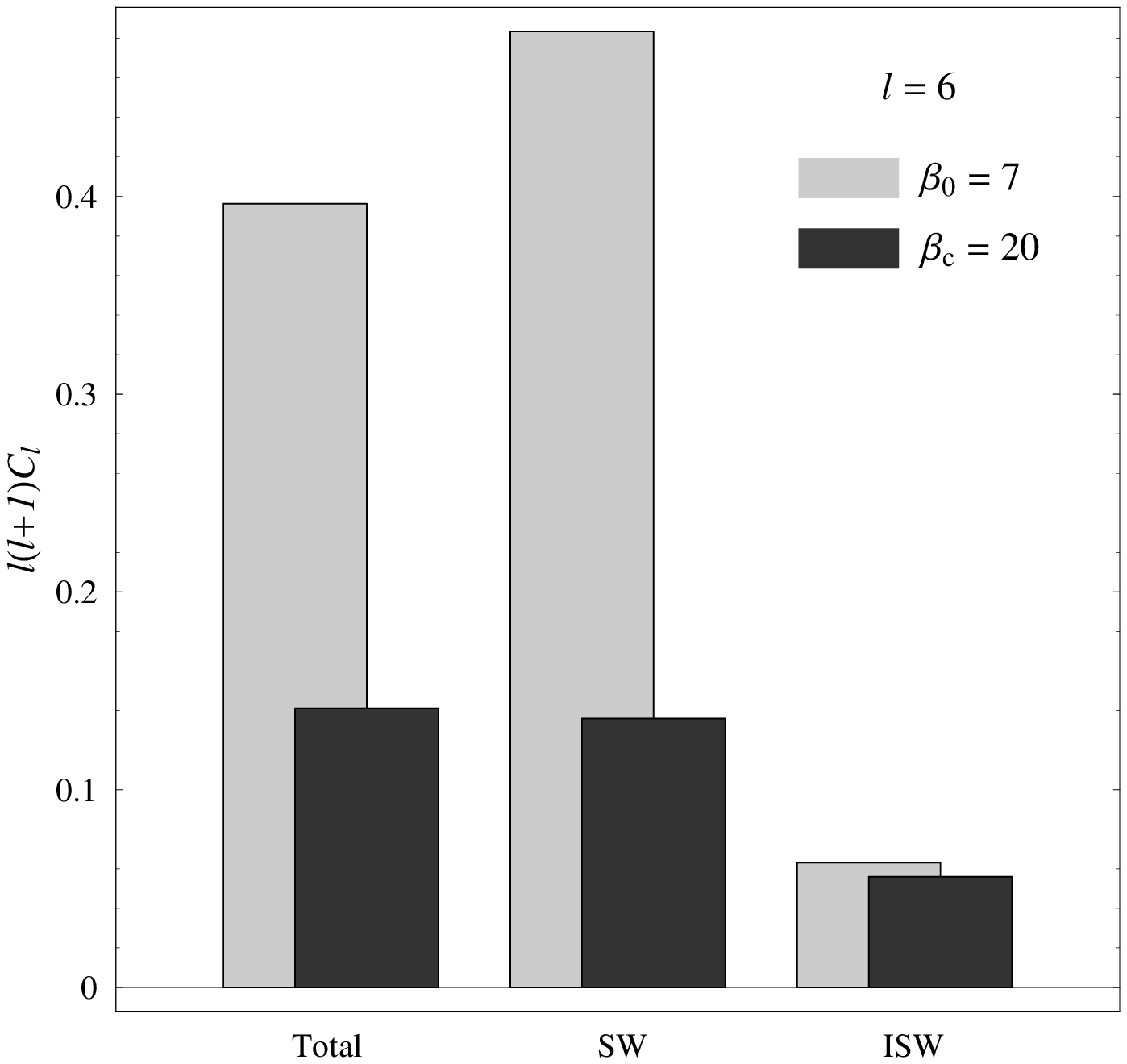}%
\hspace{0.05\textwidth}%
\includegraphics[width=0.4\textwidth]{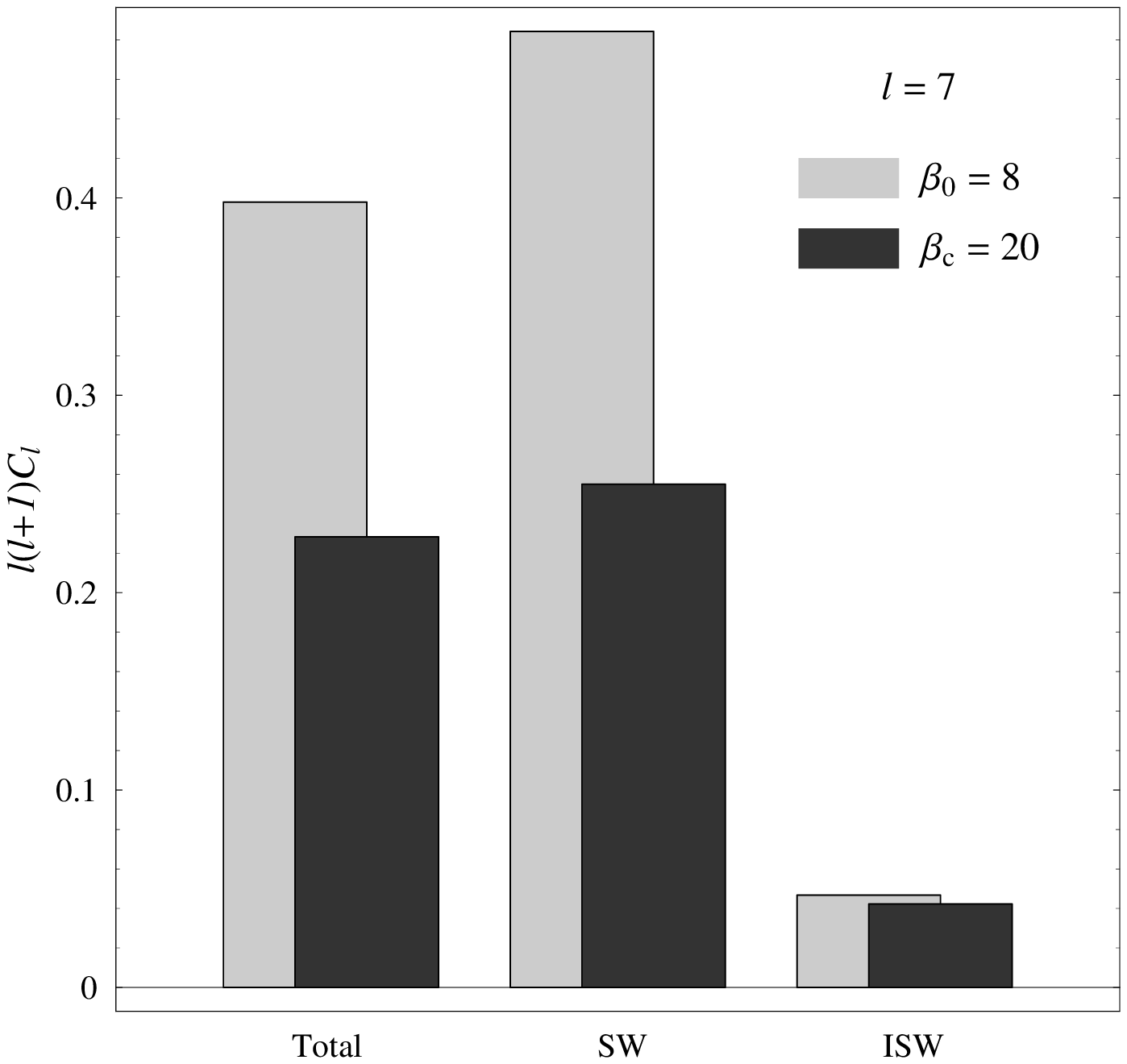}%
\end{center}
\caption{Ratios of SW and ISW contributions to low multipoles
$l=2,4,6,7$. For IR cutoff, the contributions are shown in black.
The grey is the ratio by choosing the intrinsic cutoff.}
\end{figure}

The revealed relation between the dark energy and the features at
low multipoles in the CMB power spectrum through an IR/UV duality
provides a method to constrain the equation of state of the dark
energy from the CMB data at small $\ell$, a feature we have discussed
recently \cite{recentpaper}.

We concentrate on the static equation of state for the dark energy
here.  Since the significant suppression happens at $\ell_c=2,3$, while $\ell_c>8,9$ the fluctuations is consistent with the expected value, we impose the cutoff position within the interval $3<\ell_c<7$ as a prior. 
Considering the
cutoff position within this interval, the distribution of the
corresponding $w$ for spaces with different curvature $\Omega_k^0$  got by using the cosmic duality is shown
by the solid line in Fig.2, where we have taken $\Omega_{\Lambda}^0=0.73$. The distribution is in the Gaussian form
$P(w)=\frac{1}{\sqrt{2\pi}\sigma_0}exp[-
\frac{(w-\bar{w})^2}{2\sigma_0^2}]$
where $\bar{w}=-1.19$ and $\sigma_0=0.46$; $\bar{w}=-1.15$ and
$\sigma_0=0.48$; $\bar{w}=-1.07$ and $\sigma_0=0.42$; $\bar{w}=-1.0$ and
$\sigma_0=0.43$ for $\Omega_{tot}^0=1.01, 1.02, 1.03, 1.04$,
respectively. In the $1\sigma$ 
level, $w\in[-1.65, -0.73]; [-1.64, -0.67]; [-1.49, -0.65]; [-1.43,
-0.57]$ for $\Omega_{tot}^0=1.01, 1.02, 1.03, 1.04$, respectively. 
As a comparison, in Fig.2 we have also shown the distribution of the
$w$ got by using the cosmic duality in the flat universe (dashed
line), which, using $\bar{w}=-1.26, \sigma_0=0.48$ and in the 1$\sigma$
range, leads to $w\in[-1.75, -0.78]$. The flat universe result
is consistent
when combined with external flat universe constraints including Supernova
Type Ia, CMB and Large-Scale Structure ($\omega=1.02^{+0.13}_{-0.19}$ and
$ \omega=-1.08^{+0.20}_{-0.18}$ with the use of an $\Omega_M$ prior and
from WMAP$_{ext}$ and 2dFGRS (Two Degree Field Galaxy Redshift Survey),
respectively \cite{riess}. With the current observational data,
the distribution of the $\omega$ obtained by the cosmic duality from the
closed universe is also favored. 
Thus, from the built cosmic duality, we can get information on the equation of
state from the small l suppressed CMB spectrum. Our discussion thus leads
to the fact that in the presence of spatial curvature, the central point 
of the equation of state parameter $\omega$ is closer to -1, which is 
consistent with extensive analysis implying that the equation of state is 
in the vicinity of -1. For a flat universe the distribution of $\omega$ is
very centered on phantom dark energy, a fact which we think to be at least
suspicious. These arguments mildly  favor a closed universe.

\begin{figure}[!hbtp]
\begin{center}
\includegraphics[width=0.8\textwidth]{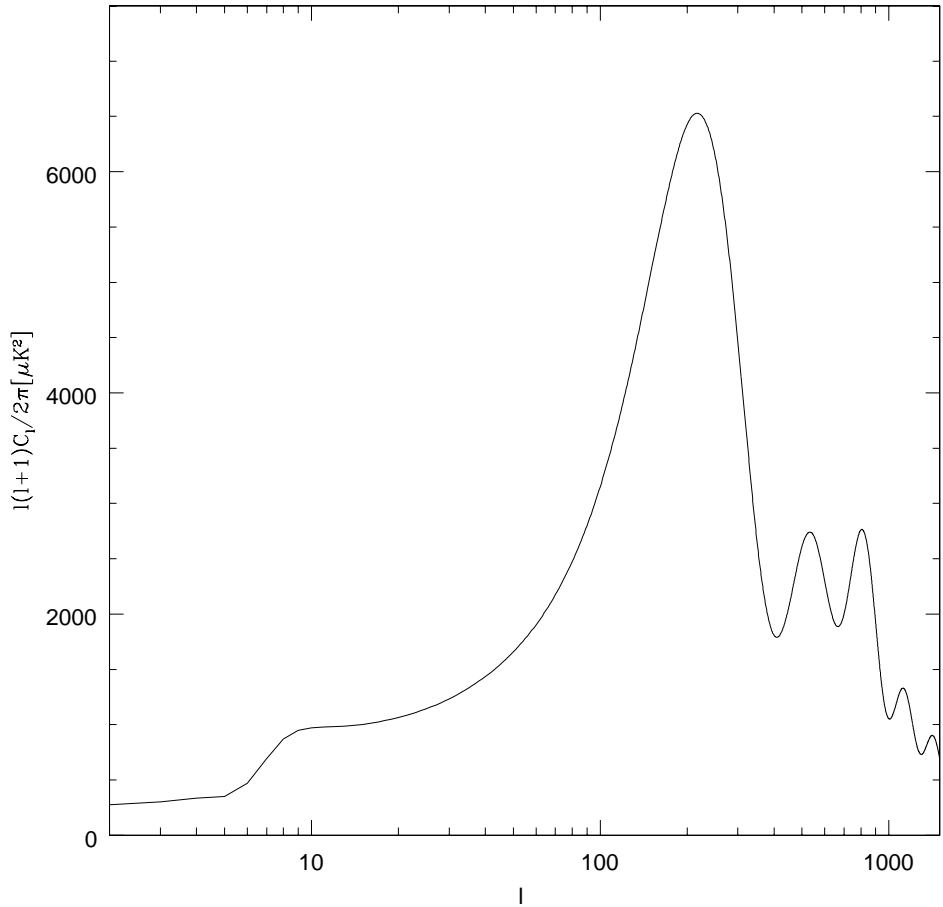}
\end{center}
\caption{CMB temperature power spectrum computed with CMBFAST with
the IR cutoff in the closed universe.}
\end{figure}

Our ability to get better measurements of the low $\ell$ power is limited
by cosmic variance. However, future observations may offer some new
insight into its origin. These observations include directly detecting the
component of the CMB fluctuations due to the ISW effect by combining the
WMAP data with traces of large scale structure, additional measurements of
fluctuations at large angular scales by observing CMB polarization and
performing TE observations to probe different regions of the sky from the
TT observations [24]. They may provide us with more accurate location of
the suppression position and in turn may give stronger constraint on the
equation of state of dark energy.  

In the study of the cosmic duality model, we see that the IR
cutoff plays an important role. In \cite{11,12}, the IR cutoff is
exactly identified with the future event horizon, $L=ar(t)$ where
$r(t)=\frac{1}{\sqrt{K}}\sin y\,$ and
$\,y=\frac{\sqrt{K}}{a}R_h=\sqrt{K}\int^{\infty}_t dt/a.$ The
holographic dark energy equation of state was derived as
%
\begin{equation}
w _0  =  - \frac{1} {3}\left( {1 + \frac{2} {d}\sqrt {\Omega
_\Lambda ^0 } \cos y} \right)\quad , \label{wandd} 
\end{equation}
where $ \cos y = \sqrt {1 - d^2 \frac{{\Omega _k^0 }} {{\Omega
_\Lambda ^0 }}} $. Then the cutoff in the wave number in the
closed universe depends on the equation of state of dark energy,
\begin{equation}
\beta _c = \sqrt {\frac{{\pi ^2 \left(
{3w_0  + 1} \right)^2 }} {{4\left( {\Omega _{tot}^0  - 1}
\right)}} + \pi ^2  + 1}
\end{equation}

Employing Eqs. (\ref{closedCl}, \ref{comovingtime},
\ref{components}) and considering that the suppression must lie within
the interval $3<l_c<7$, we obtained $w\in[-0.816, -0.534]; [-0.834,
-0.519]; [-0.833, -0.518]; [-0.867, -0.493]$ for $\Omega_{tot}^0=1.01,
1.02, 1.03, 1.04$, respectively, by 
using the idea of the cosmic duality. These results overlap with the
$1\sigma$ range of 
$w$ we got above. However, their mean values are not
within the range we obtained for $w$ by identifying the IR cutoff exactly
with the future event horizon. This shows that even if an IR/UV
duality is at work in the theory at some fundamental level, the IR
regulator might not be simply related to the future event horizon.
In \cite{17} it was argued that there might still be a complicated
relation between the dark energy and the IR cutoff of the CMB
perturbation modes. 

We treated $d$ here as a parameter.
From the correlation between the dark energy and the power
spectrum, the constant $d$ in the holographic dark energy model
\cite{11,12} has been shown to be always bigger than unity. This is an
independent support besides the requirement of satisfying the
second law of thermodynamics \cite{12}.

As in the case of the first reference in  \cite{17}, we here have not taken account of the 
integrated Sachs-Wolfe (ISW) effect either. It would be of interest to 
study the ISW effect and compare it with the SW effect. It was argued 
in \cite{28} that in the flat universe an IR cutoff can easily kill the SW
term, while not the ISW term so that the ISW contribution to the low multipole $l=2$  is not negligible compared to the SW term. 
This needs to be examined in a universe with spatial
curvature.  Including the ISW effect, eq(4) is changed into
\begin{equation} 
C_l  = A\sum\limits_{\beta  \geqslant \beta_0} {\frac{\beta } {{\beta ^2
- 1}}\mathcal{P_R} \left( \beta  \right)\left[ \frac{1}{10}{\Phi _\beta ^l
\left( \eta_0-\eta_{LS} \right)} +\int^{\eta_0}_{\eta_r} d\eta \frac{dF}{d\eta}\Phi^l_{\beta}(\eta_0-\eta) \right]^2 } \quad ,
\end{equation} 
where $F=\frac{3\Omega_m}{2}(1+z)\frac{H(z)}{H_0}\int^{\infty}_z\frac{1+z'}{(H(z)/H_0)^3}dz'$. 
$\eta_r, \eta_0$ are conformal times at recombination and the present, repectively. 
$\beta_0$ will be taken as the intrinsic cutoff or the IR cutoff in the discussion. 
For the flat case, $F$ reduces to (3) in \cite{28}. The ISW term vanishes in the matter dominated epoch.
The integrands of SW and ISW contributions look very different as shown in Fig.3 for $l=2$ and $l=7$ respectively. 
For $l=2$, as that in the flat case discussed in \cite{28}, the SW contribution is sufficiently reduced by the IR cutoff, 
while the ISW part survives. However with the increase of $l$,  we find that ISW effect will become less and less dominate. 
For l=7, ISW effect is very weak and gives way to SW effect. The ratios between ISW, SW effects and total spectrum are shown in Fig.4. 
By choosing the IR cutoff, we have shown the ratios in black. The grey is the ratio by choosing the intrinsic cutoff. 
It is clear that by choosing the IR cutoff $\beta_c=20$ at $l=7$, ISW contribution is very weak, which is only about one-tenth of the SW.
This is the reason that we do find the clear cutoff of the spectrum around l=7, where the spectrum is not contaminated by the ISW effect.  
In Fig.5, we have shown the spectrum got by CMBFAST \cite{29} by choosing the IR cutoff in the closed universe, where the ISW effect has been 
included. 
We see that the ISW effect will not change the suppression behavior of the spectrum, which agrees with the result shown in \cite{17}.

In summary, we have generalized the holographic understanding of the low
$\ell$ features in the CMB power spectrum to the case of a closed universe.
From the analysis that combined with the CMB data, we found that
the intrinsic cutoff brought by the spatial curvature of the
closed universe cannot be counted on to explain the observed
suppressions of the CMB spectrum if the suppression point $\ell_c >3$. It is
intriguing that employing the holographic cosmic duality, where
the IR cutoff plays an important role, the power spectrum for
low $\ell$ automatically fits observational data. Our result that the suppression of the CMB spectrum  happens at $\ell\sim 7$ will not be influenced by taking account of the ISW effect. The
present model shows another possibility of how holography might
affect present day cosmology.

The holographic correlation revealed between the location of the
cutoff in the CMB spectrum and the equation of state of dark
energy presents us a way to investigate the nature of the dark
energy from the CMB data at small $\ell$. Assuming that dark energy
obeys a static (namely, redshift independent) equation of state, from WMAP
data at low $\ell$, we obtained a value of $w$ consistent with other
experiments. As a
by-product, we have shown that the constant in the holographic
dark energy model \cite{11,12} is indeed bigger than unity. This
serves as an independent support for the argument, besides the use of the
second law of thermodynamics as done in \cite{12}.  The investigation
presented here shows that the holographic scheme has a possible
implication on the understanding of the dark energy. To obtain a
more precise picture of the dark energy, the exact location of the
cutoff $\ell_c$ and accurate shape of the spectrum at low $\ell$
would be crucial to use in the method discussed here.
Comparing with the observational data, the strong constraints on $w$
may help us to determine the curvature of our universe.

\begin{acknowledgments}
This work was partially supported by  NNSF of China, Ministry of
Education of China and Shanghai Education Commission. 
E. Abdalla's work was partially supported by FAPESP and CNPQ,
Brazil. R.K. Su's work was partially supported by the National
Basic Research Project of China.
\end{acknowledgments}


\end{document}